\documentclass{svproc}
\usepackage{url}

\usepackage{fancyhdr}
\usepackage{amssymb}
\usepackage{mathrsfs} 
\usepackage{amsmath}
\usepackage{color}

\def\[#1\]{$$#1$$}

\def\frac#1#2{{#1\over#2}}

\usepackage{graphicx}

\begin{document}
\mainmatter              
\title{An Overview on the Web of Clinical Data
}

\author{
Marco Gori\inst{1}\inst{2} 
}

\institute{SAILAB,
University of Siena\\
\texttt{http://sailab.diism.unisi.it},\\
\and
MAASAI, Universit\`e C\^ote d'Azur\\
\texttt{https://team.inria.fr/maasai/team-members}
}

\maketitle              

\begin{abstract}
	In the last few years there has been an impressive growth of connections between medicine 
	and  artificial intelligence (AI) that have been characterized by the specific focus on single
	problems along with corresponding clinical data. 
	This paper proposes a new perspective in which the focus is on the progressive
	accumulation of a universal repository of clinical hyperlinked data in the spirit that
	gave rise to the birth of the Web. The underlining idea is that this repository, that is
	referred to as the Web of Clinical Data (WCD), will dramatically change the 
	AI approach to medicine and its effectivness.  It is claimed that research and AI-based applications 
	will undergo an evolution process that will likely reinforce systematically the 	
	solutions  implemented in medical apps made available in the WCD.
	The distinctive architectural feature of the WCD is that this universal repository will be under control
	of clinical units and hospitals, which is claimed to be the natural context for 
	dealing with the critical issues of clinical data.
\end{abstract}

\section*{Introduction}
Eric J. Topol, in his popular book~\cite{EricTopol2011}, claims that medicine 
will inevitably be Schumpetered in the coming years. 
No matter if this excitement  is motivated, we are still missing a crucial catalyzer for strongly accelerating 
the Schumpetering. At the moment, most studies are characterized by the 
relentless search for unified standards to share data. What if we get rid of any standard?
It is worth mentioning that while from one side the diffusion of clinical data is carefully
controlled, there are initiatives that come from patients motivated by the willingness of sharing their
clinical data and experiences\footnote{See e.g.~{\tt \scriptsize https://www.patientslikeme.com/}}.
The explosion of social networks have in fact strongly favored the birth of communities 
where people like to exchange ideas, make helpful connections, and feel supported when they need it most.
On the other hand, there are also huge efforts in the building of health systems that 
can strongly benefit also from recent the development of AI, where the emphasis has been 
shifted to machine learning and processing on unstructured data. Amongst a number of attempts, the Canadian
efforts\footnote{\tt \scriptsize https://www.cifar.ca/docs/default-source/ai-reports/ai4health-report-eng-f.pdf}
is definitely worth mentioning. 

This paper promotes the idea of building a universal repository of clinical data  
to collect the health records of people,  that closely reminds the spirit of the Web. 
This universal collection would not represent only a truly paradigm shift on the access to clinical 
data by students and experts in medicine, but it would open the doors to the new grand challenge of building 
decision support systems that operate on a universal repository on the basis of the content of the health records, 
as well as with recommendations based on the discovering of similarities. 
This evolution of a such a project seems to be intimately connected with the very critical nature of clinical data.
When considering the risks, the most natural  answer is to refrain from such a crazy adventure,
but we claim that the time has come to an in-depth analysis of risks and opportunities.
The emerging novel solutions that could dramatically
improve long-term developments in medicine, which can also go beyond the borders of single countries.
Concerning the accessibility we propose that the major difference with respect to the Web,  is that only 
the Clinical Units (CU) can access, while patients are expected to benefit from new services.
Depending on specific initiatives from CU, the act of patient data donation could be rewarded in
different ways (e.g. information services related to the patient clinical condition).
The original spirit of the Web leads to conceive the traditional world-wide search engine service 
that can itself be immediately very important for the clinical activities. 
However, most important challenges are connected with the current developments of AI, 
which is expected to facilitate the development of intelligence apps. 
One can also expect that those apps will conquer the degree of autonomy that make
them nearly indistinguishable from a medical assistant operating in team under the control of 
human experts.

 An overview is given on the WCD  by focussing mostly on the technological 
side of the project. Preliminary discussions on legal issues and the business model,
along with the social implications are also given.

\section{Web of Clinical Data: The universal repository}
\label{UnivRepSec}
\begin{figure}[htbp]
		\centering
		\includegraphics[width=12cm]{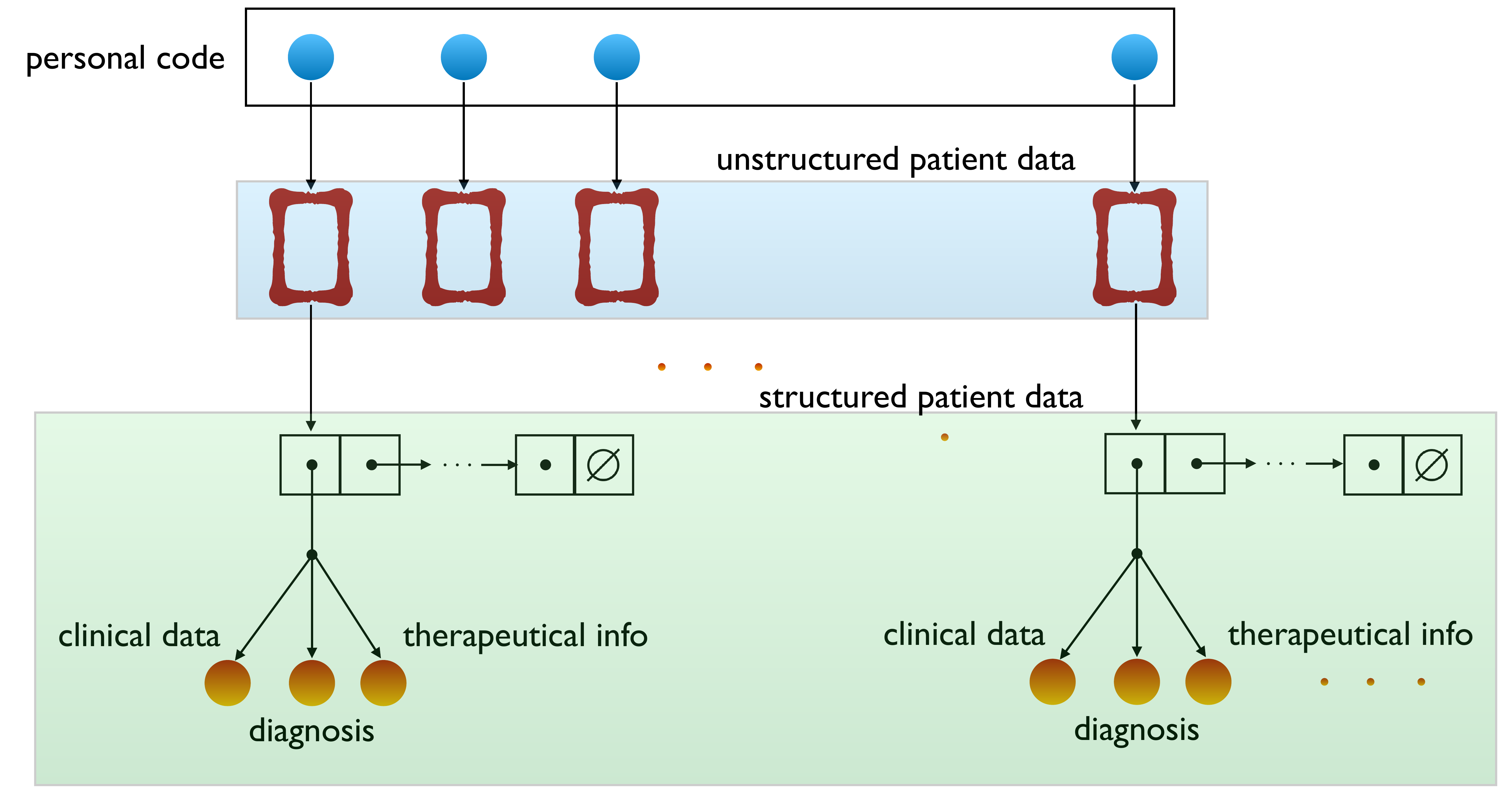}
\caption{Patients can upload their own data in a truly unstructured form. Clinical data, diagnosis,
and therapeutic documents are stored as a forest with separate data for each patient, 
that is characterized by an anonymous Personal Identification code. The separation and appropriate 
structured organization of this material can be created by opportune intelligent agents
that contribute to the creation of the WCD.}
\label{Forest-Fig}
\end{figure}

In the last few years the impressive growth of computer-based medical services
have been mostly based on appropriate organization of clinical data. 
The structured organization of huge amount of such data has been playing 
a fundamental role also in many artificial intelligence based challenges to 
medicine. Based on an early intuition~\cite{GoriWCD2014}, this paper gives a view on a
truly different way of collecting and organizing clinical data, which is claimed to
open the doors to an explosive evolution of the field.
We assume that medical information is stored in any multimedia document 
connected by hyperlinks, just like in the Web.  It's up to 
 AI agents to interpret the information hidden in the repository, which is
 referred to as the {\em Web of Clinical Data} (WCD). 
 Links in the WCD are expected to establish different levels of
relations. The nodes of the graph are medical documents that
can properly be linked and included in the WCD. Some links are used 
for connecting medical documents of a single patient corresponding to a certain 
specific clinical event, others are used for establishing relations between documents of
different patients.
No matter how the repository is created, 
we assume that documents do not contain any identifiable reference to 
the identity of patients, which are only characterized by the {\em Personal Code} (PC).
The repository can be supplied by anonymized data coming from clinical centers
as well as from patients, who upload their data after having  signed
appropriate authorizations. As a result, the repository exhibits the very simple
structure depicted in Fig.~\ref{Forest-Fig}, where any type of multimedia
medical document of a given patient is linked to his/her Personal Code. These documents 
are collected as a list for any individual thus exhibiting a forest structure. 
Ideally, one would like medical information supplied according to the
structure of  Fig.~\ref{Forest-Fig}, where the anamnesis of any patient
is represented by a list of {\em clinical events}, each of them 
properly organized into clinical data, diagnosis and therapeutic treatments. 
The distinctive underlying assumption of the WCD is that there is no need
to provide such a structure, since it is up to the  WCD Intelligent System that 
continuously process the clinical data to reconstruct the structure of Fig.~\ref{Forest-Fig}.
The identification of different clinical events might arise from the detection of the date on 
the documents. For example, it could be the case that
the therapy adopted by a patient is related to a set of clinical data and to a 
specific diagnosis, but there is no hyperlink amongst these information. Just like 
humans, it's up to an intelligent agent to realize that something is missing 
and reconstruct the links. Clearly, the presence of the date is a fundamental
cue for linking documents of the same clinical event.
However, documents with no identifiable  date can also be supplied to the WCD. 
Basically, the underlying principle is that they are also precious for medical inference, though
one must consider their reduced significance due to the lack of a temporal collocation in the
anamnesis of a patient.  

%
%

\begin{figure}[htbp]
		\centering
		\includegraphics[width=12cm]{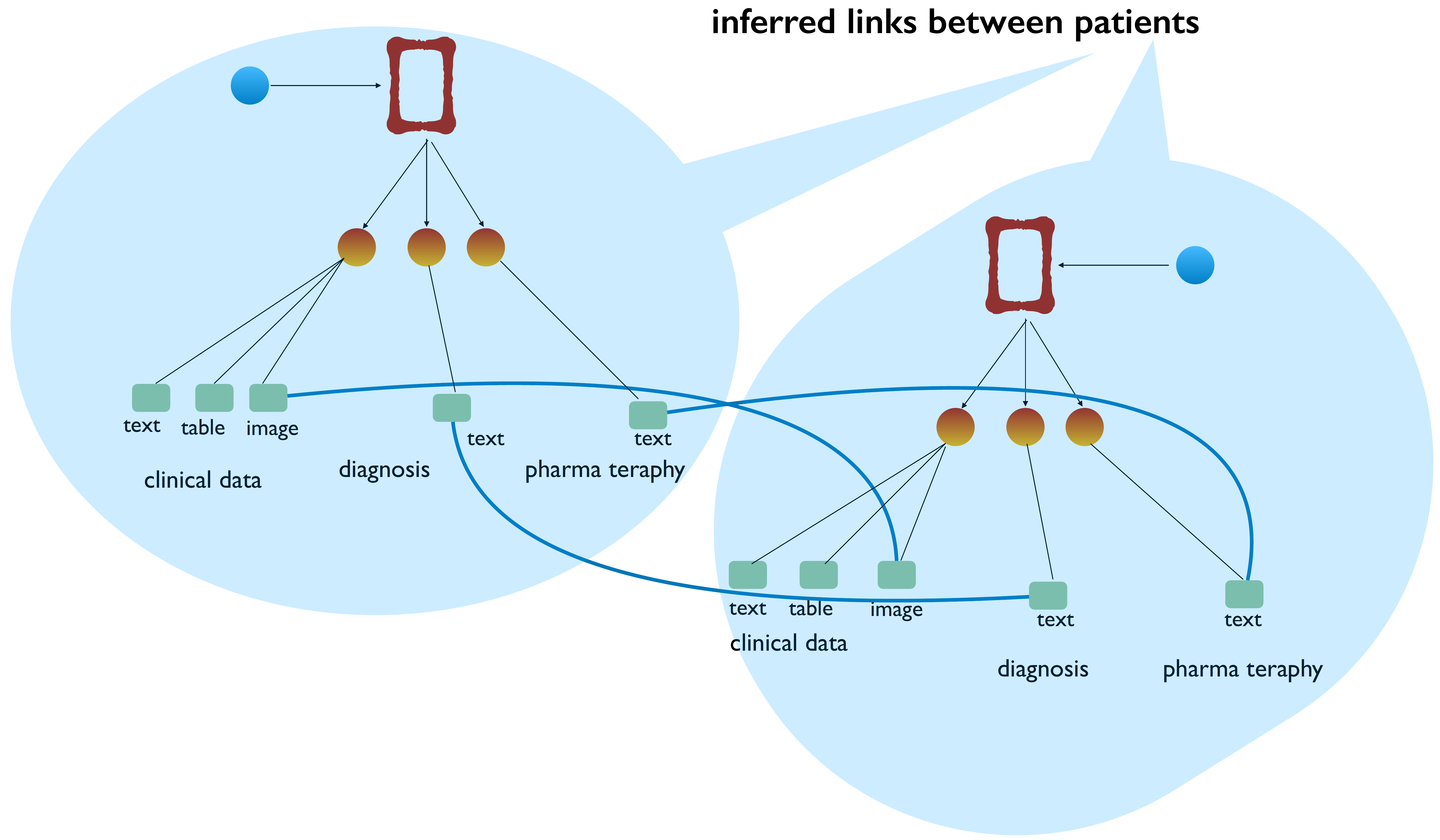}
\caption{Automatic construction of the WCD: The forest of patients grows up by 
appropriate links amongst similar data. Similarities arise from both text and
image data and are discovered by intelligent agents.}
\label{LinksBetweenPat-Fig}
\end{figure}

The role of the WCD Intelligent System goes well beyond that of reconstructing the 
structure of data of simple patients.  As we can seen in Fig.~\ref{LinksBetweenPat-Fig},
one can connect patients of the WCD so as to enrich the forest structure into a
graph-based structure where links arise because of regularities discovered in the 
repository. For example, an intelligent agent can discover similarities between the 
clinical data of two patients. This can arise from data in different formats, ranging from
text and different formats of signals. Similarities can also arise in the diagnosis and
the therapeutic treatments, so as the original forest evolves towards a truly
WCD with precious inferential information. 

\section{Data anonymization, storage, and legal issues}
The actual evolution of the WCD does require to clean a number of crucial legal issues
connected with the distribution of clinical data. Their actual distribution is supposed to 
under the control of the {\em Governing Board} of the WCD, which will be
composed of doctors, scientists in related disciplines, and layers.
First of all, we need  to clarify who is supposed to gain the permission for  accessing the data 
along with the connected purpose. The WCD is conceived for boosting research and world-wide medical 
treatment in clinical centers. As a consequence, the permission is granted to scientists and doctors
 upon approval from the {\em Governing Board} of the WCD.
As it will be shown in the next section, this restriction on the accessibility poses fundamental
constraints on the computer architecture, since the underlying assumption is that
the repository can only be made available to Clinical Units (CU), which could be hospitals,
research centers whose access comes with the duty of not to distribute data elsewhere. 
Basically, the intention is that of using the same legal framework which regulates the 
interaction of medical staff in sensible clinical data.
Following related technical developments in the field of bank accounting, we
only make the access available independently of the geographical position, whereas
we do inherit and keep all the restrictions concerning the already established 
rules for the access. Clearly, this imposes neither restrictions on the place where
data are stored nor on who provides the storing service, which does not necessarily 
corresponds with granting accessibility. 

The issue of anonymization has been the subject of an in-depth investigation
in the last few years, but only a few of them make significant efforts towards
a rigorous treatment (see e.g.~\cite{10.5555/2613665} for a very good example). 
Many circulating claims  on the issue of anonymization are often quite generic;
namely the inferential context is not well-defined, which facilitates 
restrictive claims on the supposed risks of making clinical data widely available. 
 Formally, the problem of disclosing 
the identity from medical data seems to be generic and ill-posed.
Clearly, textual documents on the anamnesis of a patient may 
contribute to discover the identity, whereas the inference from signal 
biomedical signals cannot rely on cues that can somehow reveal the identity. 
One can at most believe of some geographical connections that 
are expected to give rise to a certain disease. 
Notice that giving a collection of medical data under the conditions of the WCD,
the identity disclosure is only possible provided
that the databases of clinical centers are violated, which is in fact an
event that  can happen regardless of the WCD initiative.

Basically, the WCD initiative faces the ill-position of the issue of sensible data 
by adopting a strategy where anonymization and authorization is 
granted by design. The act of autonomously data 
uploading does exhibit the stronger patient's intention to make data
available. It is worth mentioning that the interaction with clinical 
centers follows the other way around, namely data are downloaded 
by the patients. Hence, we can think of keeping the same communication
channel where this time patients upload their own data, which can come 
from different sources. This information flow with  clinical centers 
very much resembles the one they establish with their own banks. 
The level of security and the type of information exchange shares in fact the same needs.
As patients decide to upload their data to their own preferred clinical center
(more are possible), we start creating a repository which is promoted to the
WCD, only after a further check, whose details are defined by the WCD
Governing Board.
The bottom line is that the WCD does subscribe the current scenario
for the access to clinical data, the only difference being that of making 
the data widely available to world-wide authorized clinical and research units.
In this framework, we expect that most claims on sensible information
might vanish thanks to the governing philosophy of the WCD.

\section{Overall architecture}
\begin{figure}[htbp]
		\centering
		\includegraphics[width=10cm]{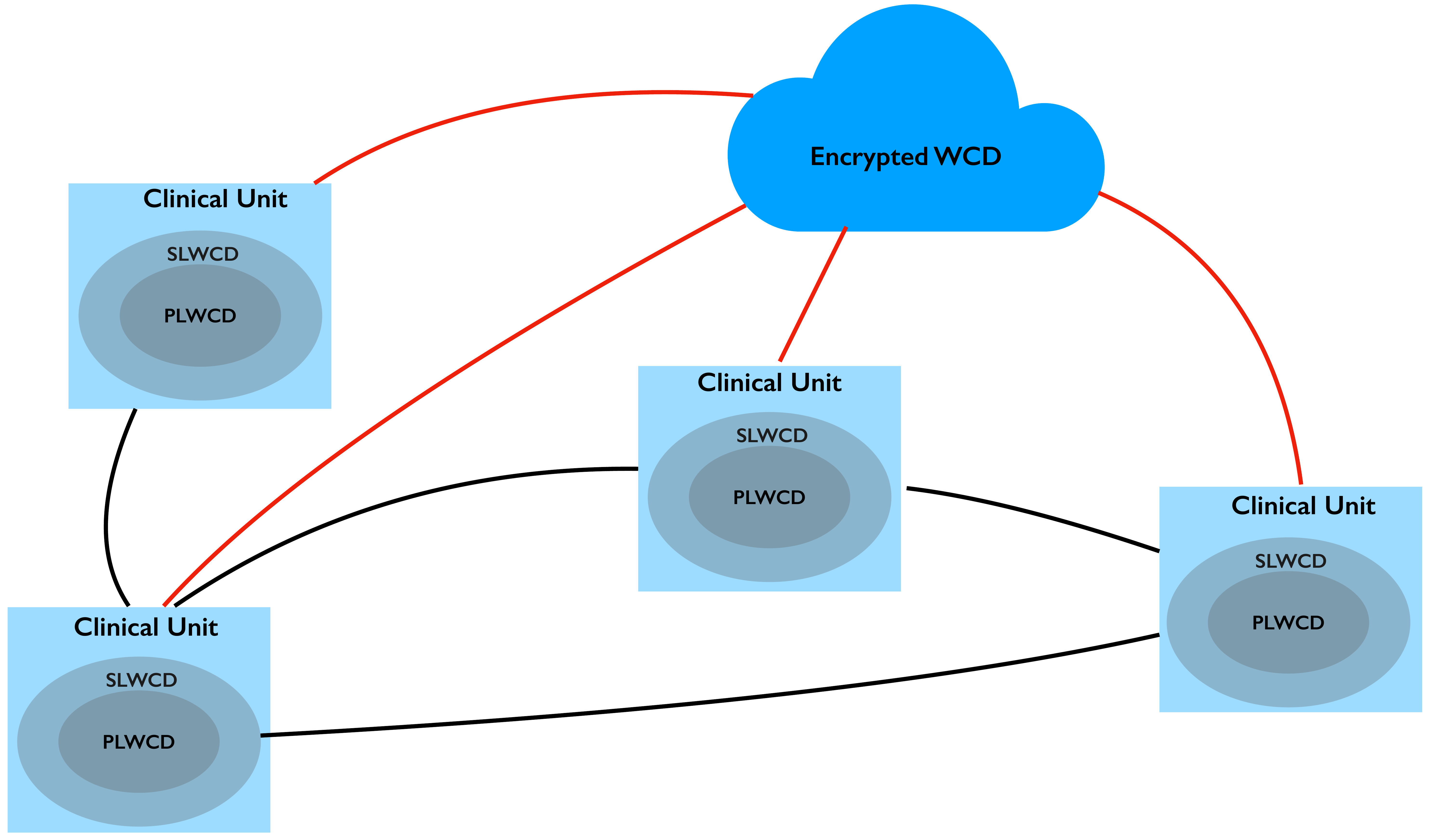}
\caption{Overall architecture of the WCD, where Clinical Units act as
macro-nodes. Each such unit is used for storing patients' data 
in a truly unstructured form (Patient Local WCD - PLWCD), while additional meta-data are
created in the (Semantic-Local WCD  - SLWCD). 
The WCD is also stored in an encrypted form onto a cloud system.  }
\label{OverallArchitecture}
\end{figure}

The access requirements pointed out in the previous sections for the WCD 
open an interesting computer architecture problem that is mostly connected with the
storing of a growing repository that must be  distributed worldwide. 
At the same time, each CU is also expected to access the whole WCD
repository.  In order to satisfy these requirement the architectural solution proposed
for the WCD is the one depicted in Fig.~\ref{OverallArchitecture}.\\
~\\
\emph{\sc WCD uniform resource locator}\\
 The repository can be given a 
formal description at local level thanks to the {\em Uniform Resource Locator}
of any document, that can be constructed by following the same principles
used in the Web. For example, the reference to the ECG diagnosis
in a certain CU could be
{\tt wcd.ao-siena.1492.ecg-careggi-june-2020.dia}, whose structure is based
on separate fields which provide information on the patient, the specific document and 
where it was archived. This is sketched below
\begin{center}
$
{\tt wcd}.\underbrace{\tt ao-siena}_{\tt Clinical Unit}.\underbrace{1492}_{\tt patient \  no}.\underbrace{\tt ecg-careggi-june-2020}
_{\tt filename}.\underbrace{\tt dia}_{\tt diagnosis}
$
\end{center}
where we easily induce the general structure of the URL.
Basically, the ECG document named {\tt ecg-careggi-june-2020.dia} is a diagnostic document
(extension {\tt dia}) corresponding to patient {\tt 1492} of Clinical Unit {\tt ao-siena} of the 
{\tt wcd}. This document is supposed to be directly accessible within the 
CU where it is stored for any type of processing. However, since this document is also supposed to 
be accessed outside the CU where it has been generated, one needs to make it available world-wide. \\
There are a number of studies on architectural solutions  for  related problems
that inspire the solution to this general framework 
(see e.g.~\cite{7349214,DBLP:journals/jisa/AlvesA18,DBLP:conf/globecom/ZittrowerZ12,10.1145/3366030.3366061}). 
Interestingly, related studies~\cite{DBLP:journals/isci/XuXLZZ20}
have also been carried out in the context of medical applications.
The solution prospected in Fig.~\ref{OverallArchitecture} is naturally following classic search engine technology
where there is a global repository. Interestingly, this analogy comes with the fundamental
difference that data are encrypted and, consequently, 
not accessible within the global Web repository.  


~\\
\emph{\sc Search engine primitive\/}\\
The basic idea is that of caching of the documents on a Cloud System (CS) in an encrypted format.
Basically, any document created in a CU is encrypted and backed up to the CS. This allows us
to carry out a classic information retrieval search~\cite{7349214} where all documents produced
in the CU are uploaded to the CS along with their encrypted keywords. This allows the CS to 
construct the inverted indexes to be used for searching. Classic encryption solutions can be 
used for handling the security of the interaction between the CUs and the CS. 
Classic  searching primitives based on propositional calculus are supposed to be used, whereas
the major assumption is not to assign the CS higher-level services. The underlying assumption
is in fact that of moving to the CUs any solution based on AI agents.  It is worth mentioning
the this horizontal keyword-based search service is in fact the first one which is offered on top
of the WCD. Interestingly, the search primitives are also of crucial importance as a building layer
for any service app.\\
~\\
\emph{\sc Service Apps\/}\\
 Any service in the WCD is supposed to be given by 
apps running in CU computer servers. Any computation does require to select the documents
that are pertinent to the service, so as they are temporarily downloaded in the CU server.
Basically, these apps are expected to operate on a proper selection of information from the WCD
that is expected to be useful for their objective. 
The structure of any service is sketched as 

\vspace{0.5cm}
\fbox{\begin{minipage}{25em}
\begin{enumerate}
	\item {\tt create the primary  app cache}
	\item {\tt q} $\leftarrow$ {\tt QueryFromService(input)} 
	\item {\tt RetrievedDocCollection}  $\leftarrow$ {\tt WCD-retrieve(q)}
	\item {\tt run BodyApp(RetrievedDocCollection,input)}
	\item {\tt free(RetrievedDocCollection)}
\end{enumerate}
\end{minipage}}
\vspace{0.5cm}

First, the app may need to create its own cache (primary cache) from the WCD to optimize
the performance. This step may also be omitted, so as all the processing
is based only on the input being process. 
The second step consists of formulating the searching query on the basis
of the input to the app. Then, as the query is fed to the {\tt QueryFromService}, 
pertinent documents are retrieved and stored in the secondary 
cache {\tt RetrievedDocCollection}. These locally retrieved collection, along with the
primary cache is used for carrying out the task assigned to the app. 
Basically, the actual processing is carried out by {\tt BodyApp} on the 
second argument ${\tt input}$ by exploiting the local cache. 
\begin{example}
	\footnotesize{
	Suppose we want to discover patients whose clinical data are similar to those
	of  a given patient. The reference for inspecting the similarity is defined by document 
	{\tt input=ecg-careggi-june-2020.dia}, which is in fact a document under diagnosis.
	This document comes with a number of attached metadata, like for instance, the specific type of 
	diagnosis (ECG signal). The document, along with its attached metadata represent
	the {\tt input} of the service, which is expected to compose the query {\tt q} 
	automatically by using {\tt QueryFromService}. As the query is obtained, it is used to
	collect {\tt RetrievedDocCollection} and, finally, the body of specific part of the app {BodyApp}
	is used to discover which documents are similar.
	}
\label{SimilarityEx}
\end{example}

\section{WCD enrichment apps}
Patients and staff from the CUs are generally expected to upload clinical data in 
the general unstructured form of a multi-media document. 
Clearly, it could be the case that single patients directly provide information
in a truly structured  way according to Fig.~\ref{Forest-Fig}.
Whenever this does not happen, the purpose of WCD enrichment apps
is that of creating the missing structured representation. 

\begin{figure}[htbp]
		\centering
		\includegraphics[width=9cm]{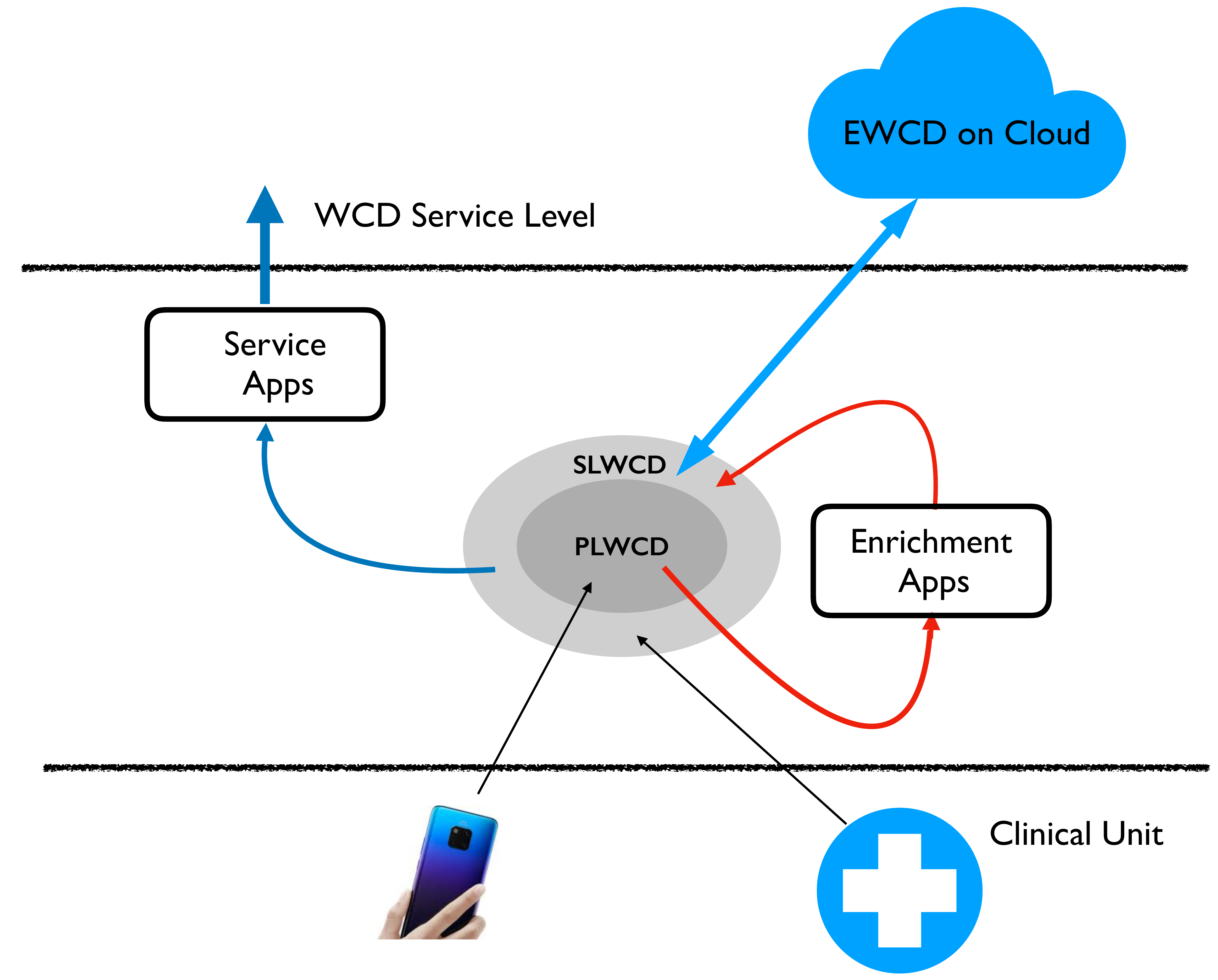}
\caption{Service and Enrichment apps in the WCD.  For a service app
to operate on the WCD the connection with the cached info in the EWCD
is required.}
\label{DAFig}
\end{figure}

The first important enrichment task is that of segmenting patients' clinical events.
The most informative cue comes from the eventual presence of the date
in the document. As it is available, the task of its detection is a well-posed
problem either for textual or image-based documents. In both cases
the extraction of the data could be quite a complex problem. However, in both
cases we are in front of classic problems of pattern recognition that have
been the subject of massive investigation (see e.g.~\cite{Stratti2008}, for early
studies). It is worth mentioning that whenever the date is either missing or badly
recognized, the segmentation of patients' event can also rely on different cues.
For example, there are cases in which a certain drug is prescribed upon 
a corresponding diagnosis, so as two documents can be placed in the 
same medical event. Clearly, this task can become very sophisticated and 
very well represents the type of challenges that are open with the WCD.

Other important enrichment tasks involve the separation between
clinic, diagnostic, and therapeutic data. This is essentially a document 
classification task. While there is a huge literature for attacking this problem
the specific context of the WCD provide a number of additional cues to 
solve it successfully.  Hence  enrichment apps contribute 
to create the structured view of the forest indicated in
Fig.~\ref{Forest-Fig}. Moreover, the successive discovery of links between
different patients of the same CU depicted in Fig.~\ref{LinksBetweenPat-Fig} 
is another tasks of enrichment apps. This problem is basically one of 
discovering the similarities between multimedia documents, which has 
also been the subject of a massive investigation 
In addition to links between documents of the same type, we can find
links between text and documents, a task which
clearly needs the additional step of extracting textual description from images.
Overall, enrichment apps create the WCD locally to each CU. However, 
their task goes beyond the discovery of local links. In order to generate global
links upon discovery of relationships between documents of different CUs,
enrichment apps rely on the software architecture described in Section~\ref{OverallArchitecture}.
In particular,  the discussion of Example~\ref{SimilarityEx} concerning  
the discovery of global  similarities for a given document provides an insight on how to attack the problem. 
Clearly, once the similarities have been found, we need to update the index of the WCD 
by its enrichment with the discovered links.
While this is carried out in the CUs, the inverted indexes in the CS must be 
updated according. Notice that the process of constructing the WCD by this similarity inspection
needs to be frequently carried out because of the continuous update of documents on the CUs.

\section{WCD service apps and medical assistants}
The creation of the WCD along with the associated high-level meta-data associated with the 
documents open the doors to the development of medical service apps ranging from 
the field of diagnosis to that of therapeutic treatment. The underlying philosophy in the 
development of service apps is that they do not simply operate on a specific CU, but
on the  WCD. Depending on the dimension of the CU, the apps might have different configurations,
but the are conceived for working at global level.

\begin{figure}[htbp]
		\centering
		\includegraphics[width=9cm]{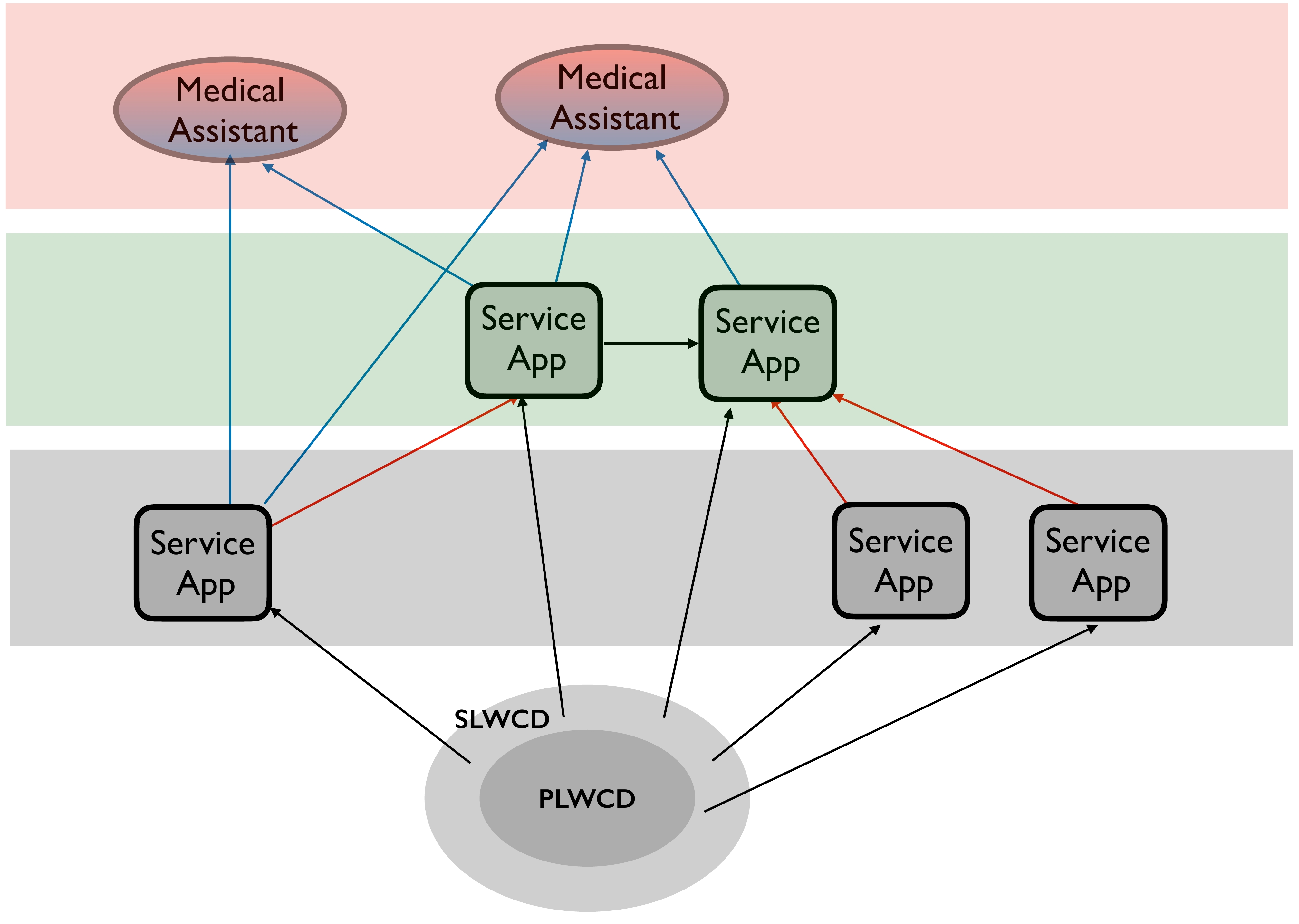}
\caption{Service apps onion skin view. The lower level 
only relies on the processing of data from the SLWCD. 
In a middle layers, apps operate by relying on the computation of
other apps. Finally, in the last layer, medical assistant operate
autonomously on the basis of all the info available in the lower levels.}
\label{SAFig}
\end{figure}

Because of the need to make them available on the WCD, 
service apps are expected to run on popular operating systems.
As we can see in Fig.~\ref{SAFig}, service apps can operate at different 
level of abstraction. As the process conquer a certain degree of autonomy,
service apps become a sort of medical assistant which can support decisions
under the control of human experts.
The long term view of the WCD is that the progressive construction of service apps
is expected to contribute to medical treatment on specific patients, but also to the
permanent creation of metadata that are expected to provide an important support 
for future medical medical decisions.

\section{Conclusions}
The described  framework of the WCD suggests the development of 
a stable marriage between medicine and artificial Intelligence. While there is a benefit in the
systematic developments of service apps thanks to the word-wide competition for medicine,
a much better benchmarking context is created for research in artificial intelligence.
Basically, it looks like the WCD can act as a fundamental catalyzer for both the disciplines.
The current focus on specific benchmarks is expected to be translated into a 
sort of permanent AI challenge in medicine, that will constantly refer to the WCD.

The philosophy behind the WCD is that of increasing the level of medical treatment by 
strongly promoting the CUs, which turn out to be the nodes of the WCD. As such, they 
are expected to control the world-wide repository of clinical data and will likely 
realize that there is a crucial benefit of assuming this role. The WCD will make available best
medical practices worldwide, a service that has an enormous value and that might also be 
paid off. Pharma companies might be interested in advertisement on the WCD, which 
could the source of huge resources especially for best CUs. Simple metrics could in fact
measure the number of access to documents uploaded to different CUs, so as 
to distribute money from advertisement. This will definitely bless the principle that in the
WCD framework doctors carry out two different specific roles: First, their primary role is
that of providing the appropriate medical assistance to single patients. Second, their
successful treatments will be useful for future worldwide medical decisions. 
Finally, the conception of the WCD comes with the wish that its development will be primarily 
useful for the medical support to poor countries. 
 

\section*{Acknowledgments}
We thank Alessandro Rossi, Stefano Melacci, Sandro Bartolini  (University of Siena),
Arturo Chiti (Humanitas, Milan), and  Paolo Traverso (FBK, Trento) for insightful discussions. 

\end{document}